\documentclass[prd,twocolumn,groupedaddress,showpacs,nofootinbib]{revtex4}
\usepackage{graphicx}
\usepackage{dcolumn}
\usepackage{amssymb}
\usepackage{mathrsfs}
\usepackage{amsmath}
\usepackage{epsfig}
\usepackage[dvips]{color}
\usepackage{hhline}

\begin{document}

\title{Radiative Dirac neutrino mass, DAMPE dark matter and leptogenesis}

\author{Pei-Hong Gu}
\email{peihong.gu@sjtu.edu.cn}

\affiliation{School of Physics and Astronomy, Shanghai Jiao Tong
University, 800 Dongchuan Road, Shanghai 200240, China}

\begin{abstract}

We explain the electron-positron excess reported by the DAMPE collaboration recently in a radiative Dirac seesaw model where a dark $U(1)_X^{}$ gauge symmetry can (i) forbid the tree-level Yukawa couplings of three right-handed neutrinos to the standard model lepton and Higgs doublets, (ii) predict the existence of three dark fermions for the gauge anomaly cancellation, (iii) mediate a testable scattering of the lightest dark fermion off the nucleons. Our model can also accommodate a successful leptogenesis to generate the cosmic baryon asymmetry.

\end{abstract}


\maketitle

\section{Introduction}

The DAMPE collaboration \cite{dampe2017} recently has reported a direct measurement of the high-energy cosmic-ray electrons and positrons with unprecedentedly high energy resolution and low background. While its largest part is well fitted by a smoothly broken power-law model, the spectrum seems to have a narrow bump above the background at around $1.4\,\textrm{TeV}$ \cite{dampe2017}. If a dark matter (DM) particle is expected to account for the DAMPE excess, it should mostly annihilate into the electron-positron pairs. Such electrophilic DM annihilation can be easily achieved in two ways: (i) a Majorana \cite{knt2003,ma2006} or Dirac \cite{bglz2009,cms2009} DM fermion couples to the charged leptons with a singly-charged mediator scalar, (ii) a DM scalar couples to the charged leptons with a singly-charged mediator lepton \cite{cehl2014}. By choosing the related Yukawa couplings, the DM could mostly couple to the electron-positron pairs and hence their annihilation could explain the DAMPE excess. There have been a number of works studying the DAMPE excess \cite{fhsty2017,gh2017,duan2017,yuan2017,fby2017,zzfyf2017,twzz2017,cy2017,gu2017-2,abfz2017,cao2017,dhwy2017,ll2017,hwzz2017,ckk2017,gm2017,niu2017}.   

On the other hand, the phenomena of neutrino oscillations have been established by the atmospheric, solar, accelerator and reactor neutrino experiments \cite{patrignani2016}. This means three flavors of neutrinos should be massive and mixed. Meanwhile, the cosmological observations have indicated that the neutrinos should be extremely light, i.e. their masses should be in a sub-eV range \cite{patrignani2016}. The tiny neutrino masses can be naturally induced at tree level by the so-called type-I \cite{minkowski1977}, type-II \cite{mw1980} and type-III \cite{flhj1989} seesaw mechanisms. Alternatively, the seesaw mechanism can be realized at loop level in association with the DM particles \cite{ma2006}. In these popular seesaw scenarios, the neutrino masses originate from some lepton-number-violating interactions and hence the neutrinos have a Majorana nature. However, we should keep in mind that the theoretical assumption of the lepton number violation and then the Majorana neutrinos have not been confirmed by any experiments. So it is worth studying the Dirac neutrinos. In analogy to the seesaw mechanisms for the Majorana neutrinos, we can construct the type-I \cite{rw1983}, type-II \cite{gh2006} and type-III \cite{gu2016} Dirac seesaw as well as the radiative Dirac seesaw \cite{gs2007} for the Dirac neutrinos.

In the Majorana or Dirac seesaw models, we can further realize a leptogenesis \cite{fy1986} mechanism with lepton number violation \cite{fy1986} or without lepton number violation \cite{dlrw1999,mp2002} in order to generate the cosmic baryon asymmetry \cite{patrignani2016}. In the Majorana or Dirac radiative seesaw models, the DM fermions can only annihilate into the lepton pairs \cite{ma2006,bglz2009,cms2009}. We hence can try to explain the DAMPE excess in these radiative seesaw models. It would be more interesting to simultaneously understand the neutrino mass, the DAMPE DM and the baryon asymmetry.

In this paper we shall consider a dark $U(1)_X^{}$ gauge symmetry to forbid the Yukawa couplings of three right-handed neutrinos to the standard model (SM) lepton and Higgs doublets. To cancel the gauge anomaly we introduce three dark fermions, which can acquire their Dirac masses with three gauge-singlet fermions after the $U(1)_X^{}$ symmetry breaking. We also impose an unbroken $U(1)_G^{}$ global symmetry under which the dark Dirac fermions, one scalar doublet and two scalar singlets are odd while the others are even. The right-handed neutrinos can obtain their Yukawa couplings to the SM lepton and Higgs doublets at one-loop level. The lightest dark Dirac fermion can keep stable to serve as a DM particle and its dominant annihilation into the electron-positron pairs can explain the DAMPE excess. Through a $U(1)$ kinetic mixing, this DM can be verified in the direct detection experiments. Furthermore, the lepton-number-conserving decays of the scalar singlets can produce a lepton asymmetry in the SM lepton doublets and an opposite lepton asymmetry in the right-handed neutrinos to realize a successful leptogenesis.

This paper is organized as follows. In Sec. II, we introduce the model. In Sec. III, we give the radiative Dirac neutrino masses. In Sec. IV, we demonstrate the leptogenesis. In Sec. V, we discuss the DM annihilations and scatterings. Finally, we make a conclusion in Sec. VI.

\section{The model}

We denote the non-SM scalars and fermions by 
\begin{eqnarray}
&&\nu_{R}^{}(1,1,0)(+1)\,,~~\chi_{R}^{}(1,1,0)(-1)\,,~~\chi_L^{}(1,1,0)(0)\,,\nonumber\\
[2mm]
&&\!\begin{array}{c}\eta(1,2,-\frac{1}{2})(0)\,,\end{array}~~\sigma(1,1,0)(0)\,,~~\xi(1,1,0)(-1)\,.~~
\end{eqnarray}
Here and thereafter the first and second brackets describe the transformation under the SM $SU(3)_c^{}\times SU(2)_L^{} \times U(1)^{}_{Y}$ gauge symmetry and the dark $U(1)_X^{}$ gauge symmetry. The right-handed neutrinos $\nu_R^{}$ and the fermion singlets $\chi_{L,R}^{c}$ carry a lepton number $L=+1$ while the scalars do not carry any lepton numbers. The model also respects a $U(1)_G^{}$ global symmetry, under which only the fermions $\chi_{L,R}^{}$ and the scalars $\eta$ and $\sigma$ are non-trivial.

We write down the Yukawa and scalar couplings relevant to the generation of DM relic, neutrino mass and baryon asymmetry, 
\begin{eqnarray}
\label{lar}
\mathcal{L}&\supset &-y_{L\alpha i}^{}\bar{l}_{L\alpha}^{} \eta \chi_{Li}^{c} - y_{Ra\alpha i}^{}\sigma\bar{\nu}_{R\alpha}^{}\chi_{Ri}^{c} +\textrm{H.c.}\nonumber\\
[2mm]
&&-f_i^{} \left(\xi\bar{\chi}_{Ri}^{}\chi_{Li}^{}+\textrm{H.c.}\right) -\mu_a^{}\left(\sigma_a^{} \eta^\dagger\phi+\textrm{H.c.}\right)\,.
\end{eqnarray}
with $l_L^{}(1,2,-\frac{1}{2})(0)$ and $\phi(1,2,-\frac{1}{2})(0)$ being the SM lepton and doublets. We also give some terms in the scalar potential,
\begin{eqnarray}
\label{kinetic}
V&\supset& \mu_\xi^2 \left(\xi^\dagger_{}\xi\right) +  \lambda_\xi^{}\left(\xi^\dagger_{}\xi\right)^2_{} + \mu_\phi^2 \left(\phi^\dagger_{}\phi\right) + \lambda_\phi^{}\left(\phi^\dagger_{}\phi\right)^2_{} \nonumber\\
&& + \lambda_{\xi\phi}^{}\xi^\dagger_{}\xi \phi^\dagger_{}\phi\,.
\end{eqnarray}
It should be noted that the $U(1)_X^{}$ gauge boson $X_\mu^{}$ has a kinetic mixing with the $U(1)_Y^{}$ gauge boson $B_\mu^{}$, i.e.
\begin{eqnarray}
\label{kinetic}
\mathcal{L}&\supset &-\frac{\epsilon}{2} B_{\mu\nu}^{}X^{\mu\nu}\,.
\end{eqnarray}

The $U(1)_G^{}$ global symmetry will not be broken at any scales. So the scalar singlets $\sigma$ and the scalar doublet $\eta$ will not develop any vacuum expectation values (VEVs). The scalar singlet $\xi$ is responsible for spontaneously breaking the $U(1)_X^{}$ gauge symmetry, i.e.
 \begin{eqnarray}
\xi =\frac{1}{\sqrt{2}}\left(v_\xi^{}+h_\xi^{}\right),
\end{eqnarray}
with $v_\xi^{}$ and $h_\xi^{}$ being the VEV and the Higgs boson.  At this stage, the fermion singlets $\chi_{L,R}^{}$ can obtain their Dirac masses,
 \begin{eqnarray}
m_{\chi_i^{}}^{}=\frac{1}{\sqrt{2}}f_i^{} v_\xi^{}\,.
\end{eqnarray}
The $U(1)_X^{}$ gauge boson $X_\mu^{}$ can also acquire its mass,
\begin{eqnarray}
m_X^2=g_X^2 v_\xi^2\,,
\end{eqnarray}
with $g_X^{}$ being the $U(1)_X^{}$ gauge coupling.

On the other hand, the SM Higgs doublet $\phi$ drives the electroweak symmetry breaking as usual,
\begin{eqnarray}
\phi=\left[\begin{array}{c}\frac{1}{\sqrt{2}}\left(v_\phi^{}+h_\phi^{}\right)\\
[2mm]
0\end{array}\right]~~\textrm{with}~~v_\phi^{}=246\,\textrm{GeV}\,.
\end{eqnarray}
The new Higgs boson $h_\xi^{}$ then can mix with the SM one $h_\phi^{}$ through their Higgs portal interaction, i.e.
\begin{eqnarray}
V &\supset& \frac{1}{2}M_{h_\xi^{}}^2 h_\xi^2 +\frac{1}{2}M_{h_\phi^{}}^2 h_\phi^2 + \lambda_{\xi\phi}^{} v_\xi^{}v_\phi^{} h_\xi^{} h_\phi^{}~~\textrm{with}\nonumber\\
&& M_{h_\xi^{}}^2=2\lambda_\xi^{}v_\xi^2\,,~~M_{h_\phi^{}}^2=2\lambda_\phi^{}v_\phi^2\,.
\end{eqnarray}
The mass eigenvalues should be 
\begin{eqnarray}
m_{h}^2&=&\frac{M_{h_\xi^{}}^2+M_{h_\phi^{}}^2-\sqrt{\left(M_{h_\xi^{}}^2-M_{h_\phi^{}}^2\right)^2_{}+4\lambda_{\xi\phi}^2v_\xi^2 v_\phi^2}}{2}\nonumber\\
&\simeq &\left(125\,\textrm{GeV}\right)^2_{}\,,\nonumber\\
M_{h'}^2&=&\frac{M_{h_\xi^{}}^2+M_{h_\phi^{}}^2+\sqrt{\left(M_{h_\xi^{}}^2-M_{h_\phi^{}}^2\right)^2_{}+4\lambda_{\xi\phi}^2v_\xi^2 v_\phi^2}}{2}\,,\nonumber\\
&&
\end{eqnarray}
corresponding to the physical states, 
\begin{eqnarray}
\left[\begin{array}{c}
h\\
[2mm]
h'
\end{array}\right] &=& \left[\begin{array}{rr}
\cos\beta&-\sin\beta\\
[2mm]
\sin\beta&\cos\beta
\end{array}\right] \left[\begin{array}{c}
h_\phi^{}\\
[2mm]
h_\xi^{}
\end{array}\right] \nonumber\\
[2mm]&&\textrm{with}~~\tan2\beta = \frac{2\lambda_{\xi\phi}^{}v_\xi^{}v_\phi^{}}{M_{h_\xi^{}}^2- M_{h_\phi^{}}^2}\,.
\end{eqnarray}

In addition, due to the $U(1)$ kinetic mixing (\ref{kinetic}), the $U(1)_X^{}$ gauge boson $X_\mu^{}$, which well approximates to a mass eigenstate for a small kinetic mixing, should be a dark photon with the couplings to the SM fermion pairs, i.e.
\begin{eqnarray}
\mathcal{L}\supset \epsilon X_\mu^{} \left(-\frac{1}{3}\bar{d}\gamma^\mu_{}d +\frac{2}{3}\bar{u}\gamma^\mu_{}u -\bar{e}\gamma^\mu_{}e \right)~~\textrm{for}~~\epsilon \ll 1\,.
\end{eqnarray}

\section{Radiative Dirac neutrino mass}

The fermion singlets $\chi_{L,R}^{}$, the scalar singlets $\sigma$ and the scalar doublet $\eta$ can mediate a one-loop diagram to give the Yukawa couplings of the right-handed neutrinos $\nu_R^{}$ to the SM lepton and Higgs doublets $l_L^{}$ and $\phi$, after the Higgs singlet $\xi$ develops its VEV to spontaneously break the $U(1)_X^{}$ gauge symmetry, i.e.  
\begin{eqnarray}
\mathcal{L}&\supset& -y_{\alpha\beta}^{} \bar{l}_{L\alpha}^{}\phi \nu_{R\beta}^{} +\textrm{H.c.} ~~\textrm{with}\nonumber\\
[2mm]
&& y_{\alpha\beta}^{} \simeq \frac{1}{16\pi^2_{}}y_{L\alpha i }^{} \frac{\mu_a^{} m_{\chi_i^{}}^{}}{M_{\sigma_a^{}}^2} y_{Ra\beta i}^\ast ~~\textrm{for}~~M_\sigma^2 \gg M_{\eta,\chi}^2\,.\nonumber\\
&&
\end{eqnarray}
It is easy to check that the effective Yukawa couplings $y_{\alpha\beta}^{}$ can be highly suppressed by the heavy masses $M_{\sigma}^{}$. This means a tiny Dirac neutrino mass.  

We should keep in mind that one massless neutrino is allowed by the experimental data. Even if the lightest fermion singlet $\chi_1^{}$ has no Yukawa couplings to the SM lepton doublets $l_L^{}$ and the right-handed neutrinos $\nu_R^{}$, we can obtain a neutrino mass matrix with two non-zero eigenvalues. If the lightest fermion singlet $\chi_1^{}$ only couples to the first generation of the lepton doublets and the right-handed neutrinos, we can choose the Yukawa couplings involving the other fermion singlets $\chi_{2,3}^{}$ to obtain a desired neutrino mass matrix, i.e 
\begin{eqnarray}
&&\frac{1}{16\pi^2_{}}y_{Le 1 }^{} \frac{\mu_a^{} m_{\chi_1^{}}^{}}{M_{\sigma_a^{}}^2} y_{Ra e1}^\ast  \nonumber\\
&&=   y_{ee}^{}-\frac{1}{16\pi^2_{}}\sum_{i=2,3}^{}y_{Le i }^{} \frac{\mu_a^{} m_{\chi_i^{}}^{}}{M_{\sigma_a^{}}^2} y_{Ra ei}^\ast\,.
\end{eqnarray}
As we will show later, by setting the following parameters, 
\begin{eqnarray} 
\label{par1}
y_{L e 1}^{2} \gg y_{L \mu 1}^{2}\,, ~y_{L \tau 1}^{2}\,,~y_{R a \alpha 1}^{2}\,,
\end{eqnarray}    
the lightest fermion singlet $\chi_1^{}$ can serve as the DM particle for the DAMPE excess. These parameter choice may be understood by additional symmetries such as the Froggatt-Nielsen mechanism \cite{fn1979}.

\section{Leptogenesis}

The scalar singlets $\sigma_a^{}$ have two decay modes,
\begin{eqnarray}
\label{decay1}
\sigma_a^{}\rightarrow \eta + \phi^\ast_{}\,,~~\sigma_a^{}\rightarrow \nu_R^{} + \chi_R^{}\,. 
\end{eqnarray} 
The subsequent decays of the scalar doublet $\eta$ are 
\begin{eqnarray}
\label{decay2}
 \eta \rightarrow l_L^{}+\chi_L^{}\,.
 \end{eqnarray}

At one-loop level, the scalar singlets $\sigma_a^{}$ can decay to generate an asymmetry stored in the scalar doublet $\eta$, an opposite asymmetry stored in the fermions $\chi_R^{}$ and an opposite lepton asymmetry stored in the right-handed neutrinos $\nu_R^{}$, i.e.
\begin{eqnarray}
\label{cpa1}
\varepsilon_a^{} &=& \frac{\Gamma(\sigma_a^{}\rightarrow \eta +\phi^\ast_{})-\Gamma(\sigma_a^{\ast}\rightarrow \eta^\ast_{} +\phi)}{\Gamma_a^{}}\nonumber\\
[2mm]
&\equiv& -\frac{\Gamma(\sigma_a^{}\rightarrow \nu_R^{} +\chi_R^{})-\Gamma(\sigma_a^{\ast}\rightarrow \nu_L^c +\chi_L^{c})}{\Gamma_a^{}}~~\textrm{with}\nonumber\\
[2mm]
\Gamma_a^{}&=& \Gamma(\sigma_a^{}\rightarrow \eta +\phi^\ast_{}) +\Gamma(\sigma_a^{}\rightarrow \nu_R^{} +\chi_R^{}) \nonumber\\
[2mm]
&\equiv& \Gamma(\sigma_a^{\ast}\rightarrow \eta^\ast_{} +\phi) +\Gamma(\sigma_a^{\ast}\rightarrow \nu_R^c +\chi_R^{c})\,.
 \end{eqnarray}
Due to the subsequent decays (\ref{decay2}), the asymmetry stored in the scalar doublet $\eta$ will be transferred to an asymmetry stored in the SM lepton doublets $l_L^{}$ and an equal asymmetry stored in the fermions $\chi_L^{}$. 
So, the decays of the scalar singlets $\sigma_a^{}$ eventually can lead to a lepton asymmetry stored in the SM lepton doublets $l_L^{}$ and an opposite lepton asymmetry stored in the right-handed neutrinos $\nu_R^{}$. As for the fermions $\chi_{L,R}^{}$, they cannot obtain any asymmetries from the above decaying processes. The lepton asymmetry stored in the right-handed neutrinos $\nu_R^{}$ do not participate in the $SU(2)$ sphalerons \cite{krs1985}, so that the lepton asymmetry stored in the SM lepton doublets $l_L^{}$ can work for a leptogenesis.

We calculate the decay width $\Gamma_a^{}$ at tree level, 
\begin{eqnarray}
\Gamma_a^{}&=& \frac{1}{16\pi^{}}\left[\left(y_R^\dagger y_R^{}\right)_{aa}^{}+\frac{2\mu_a^2}{M_{\sigma_a^{}}^2}\right]M_{\sigma_a^{}}^{}\,,
 \end{eqnarray}
and the CP asymmetry $\varepsilon_a^{}$ at one-loop level, 
\begin{eqnarray}
\varepsilon_a^{}=\frac{1}{2\pi}\sum_{b\neq a}^{} \frac{\textrm{Im}\left[\left(y_R^\dagger y_R^{}\right)_{ab}^{}\right]}{\left(y_R^\dagger y_R^{}\right)_{aa}^{}+ 2 \mu_a^2/ M_{\sigma_a^{}}^2}\frac{\mu_a^{}\mu_b^{}}{M_{\sigma_a^{}}^2-M_{\sigma_b^{}}^2}\,.
 \end{eqnarray}
The Yukawa couplings $y_{R}^{}$ can contain some CP phases. In this case, the above CP asymmetries can acquire a non-zero value. Actually, this CP asymmetry can be resonantly enhanced if the two scalar singlets $\sigma_{a,b}^{}$ have a quasi-degenerate mass spectrum \cite{fps1995}. Moreover, the decaying scalar singlets $\sigma_{a}^{}$ are gauge singlets. So we can expect a successful leptogenesis through the scalar singlet decays. More details will be studied elsewhere.

\section{DAMPE dark matter}

The lightest fermion singlet $\chi_1^{}$ can keep stable to serve as a DM particle and hence can explain the DAMPE excess if it has a mass $m_{\chi_1^{}}^{}\simeq 1.4\,\textrm{TeV}$ and mostly annihilates into the electron-positron pairs.

For the parameter choice (\ref{par1}), the fermion singlet $\chi_1^{}$ can annihilate into the first generation of the SM leptons through the $t$-channel exchange of the charged component of the scalar doublet $\eta$. The annihilation cross section is 
\begin{eqnarray}
\label{ann1}
&&\langle\sigma\left(\chi_1^{}\bar{\chi}_1^{}\rightarrow e_{L}^{}\bar{e}_{L}^{}\right)v_{\textrm{rel}}^{}\rangle =\langle\sigma\left(\chi_1^{}\bar{\chi}_1^{}\rightarrow \nu_{Le}^{}\bar{\nu}_{Le}^{}\right)v_{\textrm{rel}}^{}\rangle\nonumber\\
&& =\frac{\left|y_{Le}^{}\right|^4_{}}{16\pi}\frac{m_{\chi_1^{}}^2}{M_\eta^4}~~ \textrm{for}~~m_{\chi_1^{}}^{2}\ll M_\eta^{2}\,.
\end{eqnarray}
The dark photon $X_\mu^{}$ and the new Higgs boson $h_\xi^{}$ can mediate the $s$-channel annihilation of the fermion singlet $\chi_{1}^{}$ into the SM species. However, these annihilations can be safely ignored by taking the related masses heavy enough while the related couplings small enough. Indeed, we can fastly estimate 
\begin{eqnarray}
\label{ann2}
\langle\sigma\left(\chi_1^{}\bar{\chi}_1^{}\rightarrow h_\xi^{}\rightarrow \textrm{SM}\right)v_{\textrm{rel}}^{}\rangle &\sim& \frac{f_1^2 \lambda_{\xi\phi}^2 v_\xi^2}{4\pi M_{h_\xi^{}}^4} = \frac{ \lambda_{\xi\phi}^2 m_{\chi_1^{}}^2}{2\pi M_{h_\xi^{}}^4} \,,\nonumber\\
[2mm]
\langle\sigma\left(\chi_1^{}\bar{\chi}_1^{}\rightarrow X \rightarrow \textrm{SM}\right)v_{\textrm{rel}}^{}\rangle &\sim& \frac{g_X^2  \epsilon^2  m_{\chi_1^{}}^2 }{4\pi M_{X}^4} \,.
\end{eqnarray}

The $s$-channel contributions (\ref{ann2}) can be much smaller than the $t$-channel contribution (\ref{ann1}) for appropriate choices of the parameters. In this case the fermion singlet $\chi_1^{}$ can mostly annihilate into the electron-positron pairs to account for the DAMPE excess. At the same time, the fermion singlet $\chi_1^{}$ can scatter off the nucleons at tree level through the $t$-channel exchange of the dark photon and the mixing Higgs bosons $h_\xi^{}$ and $h_\phi^{}$. Furthermore, the Yukawa couplings in Eq. (\ref{lar}) will lead to a coupling of the dark fermion to a virtual photon and hence to the quark current. For simplicity, we only consider the mediation of the real dark photon and the virtual ordinary photon. The effective operators at quark level for the scattering should be \cite{bth2011}
\begin{eqnarray}
\mathcal{L}&\supset& a_X^{} Q_q^{}\bar{q}\gamma^\mu_{}q \bar{\chi}\gamma_\mu^{} \chi+a_\gamma^{}Q_q^{}\bar{q}\gamma^\mu_{}q \bar{\chi}\gamma_\mu^{} P_L^{}\chi ~~\textrm{with}\nonumber\\
[2mm]
&&a_X^{}= \frac{ \epsilon e g_X^{} }{M_X^2}  \,,~~a_{\gamma}^{}= \frac{e^2_{}\left|f_\alpha^{}\right|^2_{}}{16\pi^2_{} M_\eta^2} \left[\frac{1}{2} +\frac{1}{3}\ln\left(\frac{m_{\alpha}^2}{M_\eta^2}\right)\right]  \,,\nonumber\\
&&
\end{eqnarray}
with $Q_q^{}$ being the electric charge of the quark $q$. The scattering cross section can be computed by
\begin{eqnarray}
\sigma_{\chi p\rightarrow \chi p}^{}= \frac{\left(a_X^{}+\frac{1}{2}a_\gamma^{}\right)^2  \mu_r^2}{2\pi} ~~\textrm{with}~~\mu_r^{}=\frac{m_\chi^{}m_p^{}}{m_\chi^{}+m_p^{}}\,.
\end{eqnarray}

As an example, we take 
\begin{eqnarray}
&&m_{\chi_1^{}}^{}=1.4\,\textrm{TeV}\,,~~M_\eta^{}=5.9\,\textrm{TeV}\,,~~ M_{X,h_\xi^{}}^{}=3\,\textrm{TeV}\,.\nonumber\\
&&f_{Le1}^{}=\sqrt{4\pi}\,, ~~g_X^{}=1\,,~~\epsilon =0.1\,,~~\lambda_{\xi\phi}^{}\ll \epsilon g_X^{}\,,~~
\end{eqnarray}
to get  
\begin{eqnarray}
\langle\sigma\left(\chi_1^{}\bar{\chi}_1^{}\rightarrow e^{-}_{}e^{+}_{}\right)v_{\textrm{rel}}^{}\rangle\!\! &=&\!\!\langle\sigma\left(\chi_1^{}\bar{\chi}_1^{}\rightarrow \nu_{Le}^{}\bar{\nu}_{Le}^{}\right)v_{\textrm{rel}}^{}\rangle \nonumber\\
&\simeq& 2\,\textrm{pb}\,,\nonumber\\
[2mm]
\langle\sigma\left(\chi_1^{}\bar{\chi}_1^{}\rightarrow h_{\xi}^{} \rightarrow \textrm{SM} \right)v_{\textrm{rel}}^{}\rangle \!\!&\ll&\!\! \langle\sigma\left(\chi_1^{}\bar{\chi}_1^{}\rightarrow \!X\rightarrow \!\textrm{SM} \right)v_{\textrm{rel}}^{}\rangle  \nonumber\\
\!\!&\sim&\!\!7.5\times 10^{-3}_{}\,\textrm{pb}\,.
\end{eqnarray}
We hence can obtain the observed DM relic density \cite{patrignani2016} and explain the DAMPE excess. Although the dark photon has a negligible contribution to the DM annihilation, it can mediate a testable DM-nucleon scattering \cite{lcj2017}, i.e 
\begin{eqnarray}
&&a_X^{}=3.3\times 10^{-3}_{}\,\textrm{TeV}^{-2}_{}\,,~~a_\gamma^{}=-1.9\times 10^{-3}_{}\,\textrm{TeV}^{-2}_{}\,,\nonumber\\
[2mm]
&&\sigma_{\chi_1^{} p\rightarrow \chi_1^{} p}^{}\simeq 3.4 \times 10^{-46}_{}\,\textrm{cm}^2_{} \,.
\end{eqnarray}

\section{Conclusion}

In this paper we have shown the DAMPE excess can be understood together with the small neutrino mass and the cosmic baryon asymmetry in a radiative Dirac seesaw model. Specifically, the tree-level Yukawa couplings of three right-handed neutrinos to the SM lepton and Higgs doublets are exactly forbidden by the dark $U(1)_X^{}$ gauge symmetry. Then three dark fermions are introduced to cancel the gauge anomaly. When the dark Higgs singlet drives the $U(1)_X^{}$ symmetry breaking, the dark fermions can acquire their Dirac masses with three gauge-singlet fermions. The model also contains one scalar doublet and two scalar singlets. The Dirac neutrinos can obtain the tiny mass at one-loop level. The lightest dark fermion can keep stable to serve as the DM particle and can mostly annihilate into the electron-positron pairs. The $U(1)$ kinetic mixing can mediate a testable DM-nucleon scattering in the direct detection experiments. Through the lepton-number-conserving decays of the scalar singlets, the SM lepton doublets can obtain a lepton asymmetry to participate in the sphaleron processes for a successful leptogenesis, while the right-handed neutrinos can obtain an opposite lepton asymmetry.

\textbf{Acknowledgement}: P.H.G. was supported by the National Natural Science Foundation of China under Grant No. 11675100, the Recruitment Program for Young Professionals under Grant No. 15Z127060004.


\begin{thebibliography}{99}





\bibitem{dampe2017}
G. Ambrosi  {\it et al.}, (DAMPE Collaboration), doi:10.1038/nature24475.




\bibitem{knt2003}
L.M. Krauss, S. Nasri, and M. Trodden, Phys. Rev. D \textbf{67}, 085002 (2003).


\bibitem{ma2006}
E. Ma, Phys. Rev. D \textbf{73}, 077301 (2006).





\bibitem{bglz2009}
X.J. Bi, P.H. Gu, T. Li, and X. Zhang, JHEP \textbf{0904},  103 (2009).

\bibitem{cms2009} 
Q.H. Cao, E. Ma, and G. Shaughnessy, Phys. Lett. B \textbf{673}, 152 (2009).


\bibitem{cehl2014}
S. Chang, R. Edezhath, J. Hutchinson, and M. Luty, Phys. Rev. D \textbf{90}, 015011 (2014).




\bibitem{fhsty2017}
Y.Z. Fan, W.C. Huang, M. Spinrath, Y.L.S. Tsai, and Q. Yuan, arXiv:1711.10995 [hep-ph].

\bibitem{gh2017}
P.H. Gu and X.G. He, arXiv:1711.11000 [hep-ph].

\bibitem{duan2017}
G.H. Duan {\it et al.}, arXiv:1711.11012 [hep-ph].

\bibitem{yuan2017}
Q. Yuan {\it et al.}, arXiv:1711.10989 [astro-ph.HE].


\bibitem{fby2017}
K. Fang, X.J. Bi, and P.F. Yin, arXiv:1711.10996 [astro-ph.HE].




\bibitem{zzfyf2017}
L. Zu, C. Zhang, L. Feng, Q. Yuan, and Y.Z. Fan, arXiv:1711.11052 [hep-ph].

\bibitem{twzz2017}
Y.L. Tang, L. Wu, M. Zhang, and R. Zheng, arXiv:1711.11058 [hep-ph].

\bibitem{cy2017}
W. Chao and Q. Yuan, arXiv:1711.11182 [hep-ph].


\bibitem{gu2017-2}
P.H. Gu, arXiv:1711.11333 [hep-ph].


\bibitem{abfz2017}
P. Athron, C. Balazs, A. Fowlie, and Y. Zhang, arXiv:1711.11376 [hep-ph].


\bibitem{cao2017}
J. Cao  {\it et al.}, arXiv:1711.11452 [hep-ph].


\bibitem{dhwy2017}
G.H. Duan, X.G. He, L. Wu, and J.M. Yang, arXiv:1711.11563 [hep-ph].


\bibitem{ll2017}
X. Liu and Z. Liu, arXiv: 1711.11579 [hep-ph].


\bibitem{hwzz2017}
X.J. Huang, Y.L. Wu, W.H. Zhang, and Y.F. Zhou, arXiv:1712.00005 [astro-ph.HE]


\bibitem{ckk2017}
I. Cholis, T. Karwal, and M. Kamionkowski, arXiv:1712.00011[astro-ph.HE].


\bibitem{gm2017}
Y. Gao and Y.Z. Ma, arXiv:1712.00370 [astro-ph.HE].



\bibitem{niu2017}
J.S. Niu {\it et al.}, arXiv:1712.00372 [astro-ph.HE].








 
\bibitem{patrignani2016}
C. Patrignani {\it et al.}, (Particle Data Group Collaboration), Chin. Phys. C \textbf{40}, 100001 (2016).


\bibitem{minkowski1977}
P. Minkowski, Phys. Lett. B \textbf{67}, 421 (1977); T. Yanagida, in
{\it Proceedings of the Workshop on Unified Theory and the Baryon
Number of the Universe}, edited by O. Sawada and A. Sugamoto (KEK,
Tsukuba, 1979), p. 95; M. Gell-Mann, P. Ramond, and R. Slansky, in
{\it Supergravity}, edited by F. van Nieuwenhuizen and D. Freedman
(North Holland, Amsterdam, 1979), p. 315; S.L. Glashow, in {\it
Quarks and Leptons}, edited by M. L\'{e}vy {\it et al.} (Plenum, New
York, 1980), p. 707; R.N. Mohapatra and G. Senjanovi\'{c}, Phys.
Rev. Lett. \textbf{44}, 912 (1980).


\bibitem{mw1980}
M. Magg and C. Wetterich, Phys. Lett. B \textbf{94}, 61 (1980); J.
Schechter and J.W.F. Valle, Phys. Rev. D \textbf{22}, 2227 (1980);
T.P. Cheng and L.F. Li, Phys. Rev. D \textbf{22}, 2860 (1980); G.
Lazarides, Q. Shafi, and C. Wetterich, Nucl. Phys. B \textbf{181},
287 (1981); R.N. Mohapatra and G. Senjanovi\'{c}, Phys. Rev. D
\textbf{23}, 165 (1981).


\bibitem{flhj1989}
R. Foot, H. Lew, X.G. He, and G.C. Joshi, Z. Phys. C \textbf{44},
441 (1989).







\bibitem{rw1983}
M. Roncadelli and D. Wyler, Phys. Lett. B \textbf{133}, 325 (1983);
P. Roy and O. Shanker, Phys. Rev. Lett. \textbf{52}, 713 (1984).



\bibitem{gh2006}
P.H. Gu and H.J. He, JCAP \textbf{0612}, 010 (2006).



\bibitem{gu2016}
P.H. Gu, JCAP \textbf{1607}, 004 (2016).



\bibitem{gs2007}
P.H. Gu and U. Sarkar, Phys. Rev. D \textbf{77}, 105031 (2008)




\bibitem{fy1986}
M. Fukugita and T. Yanagida, Phys. Lett. B \textbf{174}, 45 (1986).


\bibitem{dlrw1999}
K. Dick, M. Lindner, M. Ratz, and D. Wright, Phys. Rev. Lett.
\textbf{84}, 4039 (2000).

\bibitem{mp2002}
H. Murayama and A. Pierce, Phys. Rev. Lett. \textbf{89}, 271601
(2002).




\bibitem{krs1985}
V.A. Kuzmin, V.A. Rubakov, and M.E. Shaposhnikov, Phys. Lett. B
\textbf{155}, 36 (1985).



\bibitem{fn1979}
C.D. Froggatt and H.B. Nielsen, Nucl. Phys. B \textbf{147}, 277 (1979).




\bibitem{fps1995}
M. Flanz, E.A. Paschos, and U. Sarkar, Phys. Lett. B \textbf{345},
248 (1995); M. Flanz, E.A. Paschos, U. Sarkar, and J. Weiss, Phys.
Lett. B \textbf{389}, 693 (1996); L. Covi, E. Roulet, and F.
Vissani, Phys. Lett. B \textbf{384}, 169 (1996); A. Pilaftsis, Phys.
Rev. D \textbf{56}, 5431 (1997).


\bibitem{bth2011}
B. Ren, K. Tsumura, and X.G. He, Phys. Rev. D \textbf{84}, 073004 (2011).


\bibitem{lcj2017}
J. Liu, X. Chen, and X. Ji, Nature Physics \textbf{13}, 212 (2017).



\end{thebibliography}
\end{document}